\documentclass[12pt]{JHEP3}

\title{Dark Energy and Cosmic Speed-Up from Consistent Modified Gravity}

\author{Shin'ichi Nojiri \\
Department of Applied Physics,
National Defence Academy, 
Hashirimizu Yokosuka 239-8686, Japan \\
email: nojiri@nda.ac.jp, snojiri@yukawa.kyoto-u.ac.jp}
\author{\speaker{Sergei D.Odintsov}\thanks{Also at TSPU, Tomsk, Russia} \\
Instituci\`o Catalana de Recerca i Estudis
Avan\c{c}ats (ICREA)  and Institut d'Estudis Espacials de Catalunya (IEEC), 
Edifici Nexus, Gran Capit\`a 2-4, 08034 Barcelona, Spain \\
email: odintsov@ieec.uab.es}
\abstract{
We review some modified gravity models  which describe
the gravitational
dark energy and the possibility of cosmic speed-up. The new consistent 
version of such theory which contains inverse and HD curvature terms
as well as new type of coupling with matter is proposed. The accelerating cosmologies are
discussed there. The structure of finite-time (sudden) singularities
is investigated.
}

\begin{document}

\tolerance=5000

\def\pp{{\, \mid \hskip -1.5mm =}}
\def\cL{{\cal L}}
\def\be{\begin{equation}}
\def\ee{\end{equation}}
\def\bea{\begin{eqnarray}}
\def\eea{\end{eqnarray}}
\def\beq{\begin{eqnarray}}
\def\eeq{\end{eqnarray}}
\def\tr{{\rm tr}\, }
\def\nn{\nonumber \\}
\def\e{{\rm e}}


The very interesting approach to resolution of dark energy problem 
which is the challenge for XXI century is related with possible 
modifications of gravity at large distances. In particular, one can consider 
gravity with the terms which are inverse on Riemannian invariants 
(Riemann and Ricci tensor, curvature). The simplest example of that sort is $1/R$ theory
\cite{CDTT} where the growth of (gravitational) dark energy is explained 
by the expansion. The inverse curvature terms may have M-theory  
or higher dimensional  origin\cite{SN,sasha}. Even, $1/R$ model may contain 
some instabilities or may not pass the solar system tests \cite{chiba},
it exists its consistent modification \cite{SNPRD,NOgrg,peng}
by higher derivative or logarithmic terms.
Moreover, its Palatini version (for recent discussion, see \cite{palatini})
may probably have less problems. Various (cosmological) aspects of modified gravity with inverse 
curvature terms have been studied in refs.\cite{mauro,worm,ANO,sergio}.
In the present report we review the modified gravity aspects which are related 
with dark energy and cosmic speed-up. The new modified gravity model 
is suggested where some cosmological solutions are given and its future singularity behaviour is discussed.  

As an interesting example the action of (large distances) modified
gravity may be taken in the following form
\be
\label{BRX1}
S={1 \over \kappa^2}\int d^4 x \sqrt{-g} \left(R - \gamma R^{-n} + L_m\right)\ .
\ee
Here $\gamma$ is  (an extremely small) coupling constant and $n$ is some
number. One can work in the original Jordan frame or, using conformal transformation
to scalar-tensor gravity , in Einstein frame (see\cite{cotsakis}).
 Without matter and when the curvature is small, 
the solution of equations of motion may be found \cite{CDTT}.
The FRW universe metric in Jordan frame is 
$ds^2= -dt^2 + a^2(t) \sum_{i=1}^3\left(dx^i\right)^2$.
The explicit FRW scale factor is 
\be
\label{XXX14b}
a\sim t^{(n+1)(2n+1) \over n+2}\ ,\quad w=-{6n^2 + 7n - 1 \over 3(n+1)(2n+1)}\ .
\ee
where $w$ is the effective equation of state parameter.
If $n>\frac{-1+\sqrt{3}}{2}$ or $-1<n<-\frac{1}{2}$, we find 
$w<-\frac{1}{3}$ and $\frac{d^2 a}{dt^2}>0$, that is, the universe is expanding with acceleration. 
The fact that gravity with inverse curvature may provide 
the gravitational dark energy model responsible for cosmic speed-up 
attracts the interest to such theory.

If $w<-1$, the universe is shrinking in the expression of $a$ (\ref{XXX14b}). If we replace the
direction of time by changing $t$ by $-t$, the universe is expanding but $t$ should be considered 
to be negative
so that the scale factor $a$ should be real. Then there appears a singularity at $t=0$,
where the scale factor $a$ diverges as $a \sim \left(  - t \right)^\frac{2}{3(w+1)}$.
One may shift the origin of the time by further changing $-t$ with $t_s - t$. Hence, in the present universe,
$t$ should be less than $t_s$ and there will appear the singularity at $t=t_s$:
$a \sim \left(t_s  - t \right)^\frac{2}{3(w+1)}$.
The future finite-time singularity is of sudden or Big Rip type.
Since $w+1=\frac{2(n+2)}{3(n+1)(2n+1)}$, it follows $w<-1$ when $-1<n<-\frac{1}{2}$. 
Here it is assumed $n>-1$, so that the Einstein term dominates  
when the curvature is small.

Let the ideal fluid is taken as the matter with the constant $w$: $p=w\rho$.
 From the energy conservation law $0={d\rho}/{dt} + 3H \left(\rho + p\right)$, 
it follows $\rho = \rho_0 a^{-3(1+w)}$ .

When the curvature is small, if $n>-1$, $f(R)$ may behave as
$f(R)\sim -\gamma R^{-n}$. In such a limit, an exact solution of the equation of motion  
is found to be \cite{ANO}
\be
\label{M8}
a=a_0 t^{h_0} \ ,\quad h_0\equiv -\frac{2n}{3(1+w)} \ ,\quad 
a_0\equiv \left[-\frac{-6\gamma h_0}{\rho_0}
\frac{\left\{\left(1+2n\right)\left(1+n\right) - (2+n)h_0\right\}}
{\left(-6h_0 + 12 h_0^2\right)^{n+1}}
\right]^{-\frac{1}{3(1+w)}}\ .
\ee
When $n=-1$, the result $h_0 = \frac{2}{3(1+w)}$ in the Einstein gravity is reproduced. 
The effective $w_{\rm eff}$ may be defined by $h_0=\frac{2}{3\left(1+w_{\rm eff}\right)}$.
By using (\ref{M8}), one finds $w_{\rm eff}=-1 + \frac{1+w}{n}$. 
Hence, if $w$ is greater than $-1$ (effective quintessence or even usual ideal fluid with positive $w$), 
when $n$ is positive, we obtain the effective phantom phase where $w_{\rm eff}$ is less than $-1$.
This is different from the case of pure modified gravity  (\ref{XXX14b}).
When $w_{\rm eff}<-1$, the curvature tends to become large. In the usual phantom model, the increase of the 
curvature generates the Big Rip type singularity (for recent discussion,
see\cite{bigrip}). In the model (\ref{BRX1}), however, when the curvature is 
large, the first Einstein-Hilbert term dominates. Hence, phantom era 
is transient.

The equation of motion may be written as:
\be
\label{BrCn5}
\frac{1}{\kappa^2}\left(R_{\mu\nu} - \frac{1}{2}g_{\mu\nu}\right) = T^{\rm dark}_{\mu\nu} + T_{\mu\nu}\ .
\ee
Here $T^{\rm dark}_{\mu\nu} \equiv \frac{\gamma}{\kappa^2} \left\{
 - nR_{\mu\nu} R^{-n-1} -  \frac{1}{2}g_{\mu\nu} R^{-n} 
+ n \left(\nabla_\mu \nabla_\nu - g_{\mu\nu} \nabla^2\right) R^{-n-1}\right\}$.
One may regard $T^{\rm dark}_{\mu\nu}$ as the gravitational dark energy. Due to the identity 
$\nabla^\mu \left(R_{\mu\nu} - \frac{1}{2}g_{\mu\nu}\right)=0 $, the total energy momentum tensor 
$T^{\rm dark}_{\mu\nu} + T_{\mu\nu}$ is conserved. If it is also assumed the conservation of the energy, the dark energy part $T^{\rm dark}_{\mu\nu}$ is also conserved. 
Then it looks that there is no flow of the energy between the matter sector and the 
dark energy one. However, it occurs via  the gravity coupling. 
In the solution (\ref{M8}), since the first Einstein-Hilbert term in (\ref{BRX1}) is neglected, 
we find $T^{\rm dark}_{\mu\nu}= - T_{\mu\nu}$, especialy, $\rho^{\rm dark}=-\rho$, that is, the variation 
of the matter energy density $\rho$ is absorbed into the dark energy density $\rho^{\rm dark}$. 
When $n<0$ and $w>0$, which gives $w_{\rm eff}<-1$ in (\ref{M8}) for small curvature, it follows
\be
\label{BrCn6}
\rho \sim -\rho^{\rm dark} \sim \rho_0 \left(t_s - t\right)^{2n} \ ,\quad 
\rho_0 \equiv 6h_0 \gamma \left( -6 h_0 + 12 h_0^2\right)^{-n-1}
\left( (2n+1)(n+1) - (n+2) h_0 \right)\ .
\ee
Comparing it with the case of the Einstein gravity,
where $\rho$ behaves as $\rho\sim t^{-2}$, $\rho$ 
 (\ref{BrCn6}) decreases more rapidly and tends to vanish at $t=t_s$. The decrease of $\rho$ generates 
the increase of $\rho^{\rm dark}$. 
Since the first Einstein-Hilbert term  (\ref{BRX1}) is neglected in the solution (\ref{M8}), we may include 
the Einstein-Hilbert term perturbatively. Using (\ref{BrCn5}), one gets 
\be
\label{BrCn7}
\rho^{\rm dark} + \rho = \frac{6h_0^2}{\kappa^2 \left(t_s -t\right)^2}
= \frac{8n^2}{3(1+w)^2 \left(t_s -t\right)^2}\ ,
\ee
which becomes large when $t$ goes to $t_s$. 
Since the usual energy density decreases when the size of the universe increases, Eq.(\ref{BrCn7}) shows 
that the dark energy $\rho^{\rm dark}$ becomes large rapidly. 
Eq.(\ref{BrCn7}) also indicates that one cannot neglect the Einstein term when $t\sim t_s$. 
When the curvature becomes large and the Einstein term dominates, the spacetime behaves as 
$a\sim t^\frac{2}{3(w+1)}$. If we perturbatively include the second $R^{-n}$ term in (\ref{BRX1}), 
by using (\ref{BrCn5})
\be
\label{BrCn8}
\rho^{\rm dark} \sim 6h_0 \left( (1 + n)(1 + 2n) - (2+n)h_0\right)\left(-6 h_0 + 12 h_0^2\right)^{-n-1}
t^{2n}\ ,
\ee
which  increases with time (transient period).

Let us now consider the following consistent modified gravity model composed of 
two theories\cite{SNPRD,dominance}: 
\be
\label{BRX1b}
S=\int d^4 x \sqrt{-g} \left\{{1 \over \kappa^2}\left(R - \gamma R^{-n}  + \zeta R^2\right) + 
\left(\frac{R}{\mu^2}\right)^\alpha L_d + L_m\right\}\ .
\ee
Here $L_d$ describes dark or usual matter coupled with gravity   and $\gamma$ and $\zeta$ 
are constants. The model admits the early time inflation and the late time dark energy 
universe in unified description like in phantom/quantum matter theory \cite{ds}.
Let $L_d$ (which also may be interpreted as part of dark energy\cite{dominance}) 
be the Lagrangian of the free massless scalar $\varphi$:
\be
\label{LR4}
L_d = - {1 \over 2}g^{\mu\nu}\partial_\mu \varphi \partial_\nu \varphi\ .
\ee
For FRW universe where $\phi$ only depends on $t$ $\left(\phi=\phi(t)\right)$ and  
$L_m=0$, the solution of 
scalar field equation is given by (with constant $q$) 
$\dot \varphi = q a^{-3} R^{-\alpha}$. 

When the curvature is small, the third $R^2$-term in (\ref{BRX1b}) becomes dominant. 
If we assume the first and the second terms could be neglected and $L_m=0$, the equation of the 
motion given by the variation over $g_{00}$ has the following solution:
\be
\label{BrCn10}
a=a_0 t^{\alpha + 2 \over 3}\quad \left(H={\alpha + 2 \over 3t}\right)\ ,\quad 
a_0^6 \equiv -\frac{\kappa^2 q^2 \left(2\alpha - 1\right)\left(\alpha - 1\right) }
{27{\mu^{2\alpha}  \left(\alpha + 2\right) \left(\alpha + 1\right)^{\alpha - 1}
\left({2 \over 3}\left(2\alpha - 1\right)\right)^{\alpha + 2}}}\ .
\ee
The effective $w$ is given by $w=-1 + {2}/({\alpha +2})$.
If $\alpha>1$ $\left(\alpha<0\right)$, the universe is accelerating (decelerating). 
On the other hand, when the curvature is small, the second $R^{-n}$ term in (\ref{BRX1b}) becomes 
dominant. The  solution is found to be
\bea
\label{BrCn12}
&& a=a_0 t^{\alpha -n \over 3}\quad \left(H={\alpha + 2 \over 3t}\right)\ ,\nn
&& a_0^6 \equiv \frac{\kappa^2 q^2 \left(2\alpha - 1\right)\left(\alpha - 1\right) }
{3{\mu^{2\alpha}  \left(\alpha - n\right) 
\left(3 + 11 n + 7n^2 - (2+n)\alpha \right)\left(\alpha + 1\right)^{\alpha - 1}
\left({2 \over 3}\left(2\alpha - 1\right)\right)^{\alpha + 2}}}\ ,
\eea
and  the effective $w$ is given by
$w=-1 + {2}/({\alpha -n})$.
Then if $\alpha - n > 3$, we have $w<-\frac{1}{3}$ and the universe is accelerating. 
Hence, for some parameters choice  
 the effective $w$ may cross $w=-1$ border: from greater than $-1$ to $w<-1$. 
This is an interesting feature of gravitational dark energy. 
When $w<-1$, there usually occurs finite-time singularity. In this model, due to 
$R^2$-term  (\ref{BRX1b}) it  does not occur eventually. 

One may consider the case that $L_m\neq 0$. We also assume the Hubble constant behaves as 
(which defines behaviour of matter from field equations) 
$H=h_0 + h_1 \left(t_s - t\right)^m$.
The equation given by the variation over $g_{00}$ in (\ref{BRX1b}) contains the terms including 
$\frac{d^2 H}{dt^2}$ but not the terms including $\frac{d^k H}{dt^k}$ $\left(k\geq 3\right)$. 
Therefore if $m>2$, the matter energy density $\rho$ should be finite:
\be
\label{BrCn15}
\rho \to \rho_0 \equiv \frac{6}{\kappa^2}\left\{ h_0^2 - \left(2+n\right)\left(12h_0^2\right)^{-n -1}\right\} 
 - \frac{q^2 \left(12h_0^2\right)^{-\alpha}(\alpha + 2)}{4\mu^{2\alpha}a_0^6}\ .
\ee
Here $a_0$ is the size of the universe at $t=t_s$. 
As $\rho>0$ usually, $\rho_0$ should be also positive. 
The equation given by the variation over $g_{ij}$ contains the terms including 
$\frac{d^3 H}{dt^3}$. Therefore if $3>m>2$, since $\frac{d^3 H}{dt^3}$ diverges, the matter pressure $p$ 
diverges:
\be
\label{BrCn16}
p \to p_0 \equiv 6h_1\left[ - \frac{1}{\kappa^2} \left\{ n(n+1)\gamma \left(12h_0\right)^{-n - 2} 
+ 2 \zeta \right\} 
 - \frac{\zeta \alpha(\alpha -1) q^2 }{2\mu^{\alpha}a_0^6 \left(12h_0^2\right)^{\alpha - 2}}
\right]m(m-1)(m-2)
\left(t_s - t\right)^{m-3} \ .
\ee
In case of the Einstein gravity with matter, such a sudden singularity appears when $0<m<1$ \cite{Barrow} 
(see also \cite{Barrow1}). 
The theory under consideration moderates the singularity.

Let us finally consider the role of $L_d$ in (\ref{BRX1b}) when it describes the scalar field $\varphi$ 
with the potential $V(\varphi)$. For simplicity, we put $\gamma=L_m=0$. When the scalar field $\varphi$ 
only depends on time, one has
\be
\label{BrCn17}
\rho_d = \frac{1}{2}{\dot\varphi}^2 + V(\varphi)\ ,\quad 
L_d=p_d = \frac{1}{2}{\dot\varphi}^2 - V(\varphi)\ .
\ee
Varying over $g_{00}$ gives:
\bea
\label{BrCn18}
0&=&-\frac{3H^2}{\kappa^2} + \frac{1}{\mu^{2\alpha}}\left[ 3\alpha 
\left(6\frac{dH}{dt} + 12 H^2 \right)^{\alpha - 1} \left(\frac{dH}{dt} + H^2 \right) p_d \right. \nn
&& \left. -3\alpha H \frac{d}{dt}\left\{\left(6\frac{dH}{dt} + 12 H^2 \right)^{\alpha - 1}p_d\right\} 
+ \frac{1}{2} \left(6\frac{dH}{dt} + 12 H^2 \right)^{\alpha}\rho_d\right]\ .
\eea
The variation over $g_{ij}$ looks like
\bea
\label{BrCn19}
0&=&\frac{1}{\kappa^2}\left(2\frac{dH}{dt} + 3H^2\right) + \frac{1}{\mu^{2\alpha}}
\left[ \left\{\left(3- \alpha\right)\frac{dH}{dt} + \left(6-3\alpha\right)H^2\right\}  
\left(6\frac{dH}{dt} + 12 H^2 \right)^{\alpha - 1}  p_d \right. \nn
&& \left. + \left(\frac{d^2}{dt^2} + 2H \frac{d}{dt}\right)
\left\{\left(6\frac{dH}{dt} + 12 H^2 \right)^{\alpha - 1}p_d\right\}\right]\ .
\eea
We now assume $H={h_0}/{t}$ $\left(\mbox{when}\ h_0>0\right)$ or 
$H={-h_0}/({t_s - t})$ $\left(\mbox{when}\ h_0<0\right)$. 
Then the solution of (\ref{BrCn18}) and (\ref{BrCn19}) is found to be
\bea
\label{BrCn21}
&& p_d=p_0 t^{2\alpha -2}\ ,\quad \rho_d = \rho_0 t^{2\alpha -2}\quad \left(h_0>0\right) \ ,\nn 
&& p_d=p_0 \left(t_s-t\right)^{2\alpha -2}\ ,\quad \rho_d = \rho_0 \left(t_s - t\right)^{2\alpha -2} 
\quad \left(h_0<0\right) \ , \nn
&& p_0 \equiv \frac{\mu^{2\alpha}\left(2 + 3h_0\right)}{\kappa^2 \left( -6h_0 + 12 h_0^2 \right)^{\alpha - 1} 
\left\{ - 3 + \alpha + \left( 6 - 3\alpha \right)h_0\right\} }\ ,\nn 
&& \rho_0 \equiv \frac{6\mu^{2\alpha} h_0 \left\{ 2\alpha + \left(3 - 6\alpha\right) h_0 
 - 6\left(1 - \alpha\right)h_0^2 \right\}}{\kappa^2 \left( -6h_0 + 12 h_0^2 \right)^\alpha  
\left\{ - 3 + \alpha + \left( 6 - 3\alpha \right)h_0\right\} }\ .
\eea
One may write
\be
\label{BrCn22}
{\dot\varphi}^2 = \rho_d + p_d = \frac{\varphi_0^2}{\alpha^2}t^{2\alpha - 2} \  \mbox{or} 
\  \frac{\varphi_0^2}{\alpha^2}\left(t_s - t\right)^{2\alpha - 2}\ , \quad 
V(\varphi) = \frac{\rho_d - p_d}{2}= V_0 t^{2\alpha - 2} \ \mbox{or} 
\ V_0 \left(t_s - t\right)^{2\alpha - 2}\ . 
\ee
Then we find
$\varphi= \varphi_0 t^\alpha$, $V(\varphi) = V_0 \left({\varphi}/{\varphi_0}
\right)^{2 - \frac{2}{\alpha}}$.
When $h_0<0$, if $\alpha\geq 1$, $\rho_d$ and $p_d$ does not diverge at $t=t_s$, which is different 
from the usual phantom matter generating the Big Rip singularity. 

Thus, modified gravity which gives very natural explanation of dark energy 
(via expansion) may unify the early time inflation and late time dark energy universe 
and may pass solar system tests. The phantom phase (if any) may be transient here 
and future finite-time singularity may be avoided or (depending from 
matter/dark energy choice) it changes the  structure.

\end{document}